\documentclass[12pt, twoside]{article}
\usepackage{amsfonts}
\newcommand{\bb}{\begin {minipage} {3cm}\begin{center}}
\newcommand{\ee}{\end{center}\end{minipage}}
\newcommand{\bc}{\begin {minipage} {2.5cm}\begin{center}}
\newcommand{\bd}{\end{center}\end{minipage}}
\newcommand {\C} {{\mathbb C}}
\newcommand {\R} {{\mathbb R}}

\newcommand {\q}{\begin{quote} \small}

\newcommand {\be}{\begin {eqnarray}}
\newcommand {\e}{\end {eqnarray}}

\begin{document}
\newtheorem {lemma}{Lemma}[subsection]
\newtheorem {theorem}{Theorem}[subsection]
\newtheorem {coro}{Corollary}[subsection]
\newtheorem {defi}{Definition}[subsection]
\newtheorem {obs}{Remark}[subsection]
\newtheorem {prop}{Proposition}[subsection]
\newtheorem {exa} {Example} [subsection]
\begin{center}
{\Large Yang-Baxter Algebra for the} \\\vspace{0.1cm}{\Large$n$-Harmonic Oscillator
Realisation of $sp(2n,\R)$}\\
\vspace{1cm}

\begin{sl}
{\large A.J. Macfarlane and F. Wagner}\\
D.A.M.T.P\\
Silver Street, Cambridge CB3 9EW, UK \\
\end{sl}
\vspace{0.5cm}
\today
\end{center}
\vspace{1cm}
\begin{abstract}
Using a rational $R$-matrix associated with the $4\times4$ defining
matrix representation of $c_2 \cong sp(4)$, the Lie algebra of
$Sp(4)$, a one-site operator solution of the associated Yang-Baxter
algebra acting in the Fock space of two harmonic oscillators is
derived. This is used to define $N$-site integrable systems, which are soluble
by a version of the algebraic Bethe ansatz method without nesting. All
essential aspects of the work generalise directly from $c_2$
to $c_n$. 
\end{abstract}

\section{Introduction}
One can see from references such as \cite{IKMT} that the algebraic Bethe
ansatz \cite{FAD} continues to be a subject of interesting  development and
improvement. This paper is devoted to applying the method to integrable models
with underlying Lie algebras other than $a_1 \cong su(2)$ in a way that 
produces {\it unnested} Bethe equations.

Let ${\cal V}$ and ${\cal H}$ denote auxiliary and quantum spaces respectively.
Define $R(u)$ with spectral parameter $u$ in ${\cal V} \otimes {\cal V}$, and
$T(u)$ in ${\cal V} \otimes {\cal H}$. Then the Yang-Baxter equation for $R(u)$
on ${\cal V} \otimes {\cal V} \otimes {\cal V}$,
\be\label{1.1}
R_{12}(u-v) R_{13}(u)R_{13}(v) = R_{23}(v) R_{13}(u) R_{12}(u-v) \quad ,
\e
\noindent and the Yang-Baxter operator algebra relation
on ${\cal V} \otimes {\cal V} \otimes {\cal H}$,
\be\label{1.2}
R_{12}(u-v) T_1 (u) T_2(v) = T_2(v) T_1(u) R_{12}(u-v) \quad ,
\e\noindent are objects of central importance to the study of integrable 
systems. In contexts in which there is an underlying algebra of type $a_n$,
especially $a_1$, things are quite well understood. See {\it e.g.}
\cite{jimbo,skly,GraS}. However, even for $a_n$ for $n>1$,
solution of related models by the Bethe ansatz method is iterative, giving
eigenstates specified by means of nested Bethe equations \cite{Resh1}. The same
applies also to models with $so(n)$ or $c_n \cong sp(2n)$ as underlying Lie 
algebra \cite{Resh2,Resh3}. A paper that describes the nested Bethe 
ansatz method clearly is \cite{deV}. Papers which state the case for seeking,
and which implement, alternatives to that method with its evident
complications are \cite{FFR} (on Gaudin models) and \cite{BABfl}.

Here we deal with theories governed by $c_n \cong sp(2n)$. We first obtain a
solution of (\ref{1.1}), and then a new `one-site' solution
$L(u)=T(u)$ of (\ref{1.2}) in ${\cal V} \otimes {\cal H}$ when ${\cal H}$
is the representation space of the metaplectic representation ${\cal M}_n$ of
$c_n$ defined in terms of $2n$ pairs of harmonic oscillator variables. This 
is the main result of the paper. Given this $L(u)$ however, we can proceed
directly down the path laid down by Sklyanin \cite{skly} for the
construction of families of $N$-site integrable models.

The integrable models just indicated are of interest because they are soluble
by means of a one stage application of the Bethe ansatz method, giving rise to
Bethe equations without any nesting, equations that are no more complicated 
than those familiar within $a_1$ studies \cite{GraS}.

In this paper, we present some essential results for $c_2$ and indicate their 
extension to $c_n$ in section 2, describe  ${\cal M}_n$ in section 3, and
derive our formulas for $R(u)$ and $L(u)$ in section 4. Section five contains a
sketch of the solution of our $c_2$ integrable models by our Bethe ansatz
method. Fuller discussion of this last topic and various related matters
will be provided in papers in preparation 
\cite{MPW}.

\section{The Lie algebra $c_n$}

We present the basis of $c_2$ in Cartan-Weyl form with simple
roots ${\bf r}_1=(1,-1), {\bf r}_2=(0,2)$, so that the other positive roots are
${\bf r}_3={\bf r}_1 +{\bf r}_2=(1,1)$ and ${\bf r}_4 =
{\bf r}_1+{\bf r}_3=(2,0)$.Let ${\bf H} = (H_1,H_2)$ denote
the generators of the Cartan subalgebra of $c_2$. Let 
$E_{\pm \alpha}, \alpha \in \{1,..,4\}$, denote
the raising and
lowering generators. Then the Lie algebra $c_2$ is specified by
\begin{eqnarray}\label{2.1}
[H_i, E_{\pm \alpha}]=\pm ({\bf r}_{\alpha})_i E_{\pm \alpha} & , & \quad
[E_{\alpha}, E_{-\alpha}] ={\bf r}_{\alpha}.{\bf H}  \quad , \nonumber \\
{[}E_1 , E_2{]} =\sqrt{2} E_3 & , & \quad [E_1,E_3]=\sqrt{2} E_4 \; , \quad 
\mbox{etc.} 
\end{eqnarray}

We also use a Cartesian basis $X_i, i \in \{1, \dots ,10\}$, for $c_2$
\be\label{2.2}
H_i=X_1 \; , \; H_2=X_2 \; \ \; \sqrt{2} E_{\pm \alpha}=X_{2\alpha+1}
+iX_{2\alpha+2} \; .
\e\noindent Then we may define the structure constants of $c_2$ via
\be\label{2.3}
[X_i ,X_j]=ic_{ijk} X_k \quad ,
\e\noindent and the quadratic Casimir is
\be\label{2.4}
{\cal C}^{(2)}=X_i X_i=H_1^2+H_2^2+\sum_{\alpha} \{ E_{\alpha}\, , \, 
E_{-\alpha}\} \quad .
\e
We also require the matrices of the four-by-four defining representation
${\cal D}$ of $c_2$, defined according to
\be\label{2.5}
X_i \mapsto x_i \quad ; \; H_1 \mapsto h_1 \; , \;
H_2 \mapsto h_2 , \quad E_{\pm \alpha} \mapsto e_{\pm \alpha} \; .
\e
We can display these all at once by use of a
$4\times 4$ matrix $C= x_i \otimes X_i \equiv x_i X_i$ that is very useful for 
work on the $c_2$ Yang-Baxter algebra. We have
\be\label{2.6}
C &=& H_1 h_1 + H_2 h_2 + \sum_{\alpha=1}^4 (E_{\alpha} e_{-\alpha} +
E_{-\alpha} e_{\alpha})\nonumber\\
&=& \left(\begin{array}{cccc} H_1 & E_{-1}& E_{-3} & \sqrt{2} E_{-4}\\
E_1 & H_2 & \sqrt{2} E_{-2} & E_{-3}\\
E_3 &\sqrt{2}E_2 & -H_2 & -E_{-1}\\
\sqrt{2} E_4 &E_3 & -E_1 & -H_1\end{array} \right) \quad . 
\e
One may read the explicit forms for $h_1,h_2,e_{\pm\alpha}$ off (\ref{2.6}),
verify that they do represent (\ref{2.1}) correctly, and note the hermiticity
properties
$h_1^{\dagger}=h_1$, $h_2^{\dagger}=h_1$,
$e_{\pm\alpha}^{\dagger} = e_{\mp \alpha}$ 
appropriate to generators of the compact representation ${\cal D}$ of $c_2$.
It can further be seen that the matrices
$x_i, i \in \{1,2...10\}$, possess the properties
\be\label{2.7}
x_i^{\dagger} = x_i, \quad {\mbox{Tr}} \,(x_i) = 0, \quad 
{\mbox{Tr}} \, (x_i x_j) = 2
\delta_{ij}, \quad x_i^{T} = - J x_i J^T \quad .
\e
Here $J$, the standard $c_2$ symplectic form, in our
representation, has, as its only non-zero elements: $J_{14}=J_{23}
=-J_{32}=-J_{41}=1$. In view of (\ref{2.7}), it follows that the matrices
$x_i$ satisfy the completeness relation
\be
{x_i}_{ac} {x_i}_{bd} = \delta_{ad} \delta_{bc} - J_{ab}J_{cd} \quad 
{\mbox{or}} \quad
x_i \otimes x_i = P-K \quad, \label{2.8} 
\e
where $K_{ab,cd}=J_{ab}J_{cd}$, is proportional to the projector onto the 
symplectic trace, 
and $P$ is the permutation map  on $\C^4 \otimes \C^4$. 

In the defining representation $X_i \mapsto x_i$ the
Casimir ${\cal
C}^{(2)}$ has eigenvalue $5$. In the adjoint representation
$(X_i)_{jk}=-ic_{ijk}$, it has eigenvalue $12$, so that also
\be\label{2.9}
c_{ijk}c_{ijl} = 12 \delta_{kl} \quad .
\e
Our work on the Yang-Baxter algebra requires us to extend the span of 
the $x_i$ to the
space of all traceless hermitian $4\times 4$ matrices. Thus we define
a specific set of five matrices $y_a$, $a \in \{1,2,..5\}$,
with the properties
\be y_a^{\dagger} = y_a,\quad \mbox{Tr} \,  y_a = 0,\quad \mbox{Tr} \, 
y_a y_b = 2\delta_{ab}, \quad y_a^T = J y_a J^T \quad , \label{2.10}
\e
so that the $Jx_i$ define a set of 10 symmetric matrices, whereas $Jy_a$
and $J$ itself define a set of $6$ antisymmetric ones. Further we can write
\be\label{2.11}
x_i x_j + x_j x_i = \delta_{ij} + 2 d_{ija} y_a \quad ,
\e
which defines $d_{ija} = d_{jia}$ such that $d_{iia}=0$. Use of the
completeness relation (\ref{2.7}) and trace properties, allows
proof of the results
\be\label{2.12}
d_{ija} d_{ijb} = 3 \delta_{ab} \quad , \quad y_a = {\textstyle{\frac{1}{3}}} 
d_{ija} x_i x_j \quad .
\e
Our definition (\ref{2.12}) of the $y_a$ reflects the fact that the
set $\lambda_A = \{x_1, ..., x_{10}, y_1, ..., y_5\}$ can serve as a
set of 15 $a_3$ Gell-Mann $\lambda$-matrices which satisfy
\be\label{2.13}
\lambda_A^{\dagger} = \lambda_A, \quad \mbox{Tr} \, \lambda_A =0, \quad
\mbox{Tr} \, \lambda_A \lambda_B = 2 \delta_{AB}, \quad \lambda_A \otimes
\lambda_A = 2P -{\textstyle{\frac{1}{2}}} I_4 \quad ,
\e where $I_4$ is the $4 \times 4$ unit matrix.
If we write $X=x_i \otimes x_i$, $Y=y_a \otimes y_a$, then the last
part of (\ref{2.13}) and (\ref{2.8}) read as
\be\label{2.14}
X+Y = 2P-{\textstyle{\frac{1}{2}}} I_4 \; , \quad X = P-K \quad ,
\e
which can be solved for $P$ and $K$ in terms of $I_4 \; ,\; X \; , \; Y$.

\section{The Metaplectic Representation, ${\cal M}_n$ of $c_n$.}

Just as the metaplectic representation of $c_1 \cong a_1$ is discussed
using the Fock space of one harmonic oscillator \cite{Itz,SW}
so also is
that of $c_n$ discussed using $n$ independent oscillators. We present
results explicitly for $c_2$ in a form that generalises completely and
naturally for $c_n$ for all $n$.

Since we define the row vector $v^T$ in terms of two sets of standard
harmonic oscillator variables by
\be\label{3.1}
v^T = (a_1^{\dagger}\; a_2^{\dagger} \;a_2 \;a_1) \quad ,
\e
then, using also the definition of $J$ given above,
it can be shown explicitly that the operators
\be\label{3.2}
X_i = {\textstyle{\frac{1}{2}}} v^T x_i J v
\e
obey the same commutation relations (\ref{2.3}) as the matrices $x_i$ of
(\ref{2.5}). Then (\ref{2.2}) leads us to the explicit results
\be\label{3.3}
H_1 = {\textstyle{}}\frac{1}{2} \{a_1, a_1^{\dagger}\}, \quad H_2 = 
{\textstyle{\frac{1}{2}}} \{a_2,
a_2^{\dagger}\} & , & \quad
E_{+1} = a_1^{\dagger} a_2, \quad E_{-1} = a_1 a_2^{\dagger} \, ,
\nonumber\\
E_{+2} = - a_2^{\dagger 2} /\sqrt{2}, \qquad E_{+3}&=& -a^{\dagger}_1
a_2^{\dagger},\qquad E_{+4} = -a_1^{\dagger 2}/\sqrt{2} \, , \\
E_{-2} = a^2_2/\sqrt{2} , \qquad E_{-3}&=& a_1 a_2, \qquad E_{-4} = 
a_1^2/\sqrt{2}
\, . \nonumber
\e
One can also check directly that (\ref{3.3}) correctly provides a
representation of (\ref{2.1}). From (\ref{3.3}), we find the hermiticity
conditions $H_1^{\dagger} = H_1 , \quad H_2^{\dagger} = H_2 , \quad 
E_{+1}^{\dagger} =E_{-1} , \quad  E_{\alpha}^{\dagger} = - E_{-\alpha} , \; 
\alpha \in \{ 2,3,4 \}$. These reflect the fact that (\ref{3.3}) generates the
infinite dimensional unitary metaplectic representation ${\cal M}_2$ of the 
real non-compact groups $Sp(2,\R)$.
Also the maximal compact subalgebra of (\ref{3.3}) is that of an
$SU(2)\times U(1)$ group, with angular momentum type generators given
by
\be\label{3.4}
J_z ={\textstyle{\frac{1}{2}}} (H_1 - H_2), \quad J_+ = a^{\dagger}_1 a_2 = 
E_{+1} = J_-^{\dagger}, \quad J_- = a_2^{\dagger} a_1 = E_{-1}
\e
and $U(1)$ generator $\lambda = \frac{1}{2} (H_1 + H_2)$, $[\lambda,
\bf{\vec{J}}] = 0$. In addition, we note the two commuting $a_1$ subalgebras
generated by $H_1, E_{\pm 4}$ and $H_2, E_{\pm 2}$. These correspond
to the two $c_1$ metaplectic subrepresentations of our $c_2$
metaplectic representation (\ref{3.3}).

We  also need to verify that the `non-compact' raising and lowering 
generators of $c_2$ can
be arranged to define vector operators $\vec{\xi}$ and $\vec{\chi}$,
using the standard Racah definition of tensor operators. Explicitly
$\vec{\xi}$ and $\vec{\chi}$ have spherical components
\be\label{3.5}
(\xi_{-1}, \xi_0, \xi_{+1})= (E_2, E_3, E_4) \quad , \quad
(\chi_{-1}, \chi_0,\chi_{+1}) = (E_{-4}, -E_{-3}, E_{-2}) \quad .
\e
All the required properties can be checked explicitly. It is significant
for the Bethe ansatz solution of the integrable models we construct,
that the components of the vectors $\bf{\xi}$ commute with each other,
as do those of $\bf{\chi}$.

We are now in a position to derive the key algebraic result for
${\cal M}_2$, a result which allows us to obtain an
infinite dimensional operator solution of (\ref{1.1}) when ${\cal M}_2$ is 
taken as the quantum space ${\cal H}$,
(and similiarly for $c_n$). To this end we use (\ref{3.2}) and (\ref{2.8})
to write the matrix $C=x_i X_i$ of (\ref{2.2}) in the form
\be\label{3.6} 
C = -{\textstyle{\frac{1}{2}}} I_4 + (Jv)v^T \quad ,
\e
whence it follows that $C$ obeys the quadratic relation
\be\label{3.7}
C^2 = -3C -{\textstyle{\frac{5}{4}}} I_4 \quad .
\e
We stress that this non-trivial result holds for the
particular (metaplectic) representation (\ref{3.3}) of $c_2$. 
It does not hold in an arbitrary representation of $c_2$. 

Since $\mbox{Tr} \, C=0$, (\ref{3.7}) implies that for ${\cal M}_2$
we have ${\cal C}^{(2)}={\textstyle{\frac{1}{2}}} \mbox{Tr} \, C^2=
  -{\textstyle{\frac{5}{2}}}$, a result that can be checked
directly to be correct by inserting (\ref{3.3})
into (\ref{2.4}), and expressing everything in terms of
$N_1=a_1^{\dagger} a_1$, $N_2=a_2^{\dagger} a_2$.

The entire discussion generalises readily from $c_2$ to $c_n$. The 
generalisation of (\ref{3.2}), and of the completeness relations (\ref{2.10})
are obvious, and imply that the only change needed in (\ref{3.6}) replaces the
unit matrix $I_4$ of $\C^4$ by $I_{2n}$.
Then we have
\be\label{3.8}
C^2 =- (n+1) C -\left({\textstyle{\frac{2n+1}{4}}}\right) I_{2n} \quad ,
\e
so that
\be\label{3.9}
{\cal C}^{(2)} = {\textstyle{\frac{1}{2}}} {\mbox{Tr}} \, C^2=
-{\textstyle{\frac{1}{4}}} \mbox{dim} (c_n).
\e

\section{The Yang-Baxter Equation and Algebra of $c_n$}

Let ${\cal V}=\C^4$. Then we seek a solution of
the Yang-Baxter equation (\ref{1.1}) of $c_2$ of the form
$R(u) = u I + \eta P + C(u) K$, or
\be\label{4.1}
R_{ab,cd}(u) = u \delta_{ac}\delta_{bd} + \eta \delta_{ad}\delta_{cb} +
C(u) J_{ab}J_{cd} \quad .
\e
This is a natural generalisation of the work for $a_{2n}$ where only $I$
and $P$ are invariant under the action of $sl(2n,\C)$, to the present
case where $K$ also is invariant under the action of $sp(2n,\C)$.

Equation (\ref{4.1}) yields a solution of (\ref{1.1}) provided that 
$C(u)$ is given by
\be\label{4.2}
C(u) = \frac{u\eta}{\lambda - u}, \quad \lambda = -3 \eta \quad ,
\e
and the same holds for $c_n$ when $\lambda = -(1+n)\eta$. These
results are the symplectic analogues of results found for the
orthogonal groups, {\it e.g.} in \cite{KS}, and agree with results
presented in different form, {\it e.g.} in \cite{jimbo}. We wish to use the
$R$-matrix (\ref{4.1}), (\ref{4.2}) to reach solutions of the
Yang-Baxter algebra on ${\cal V} \otimes {\cal V} \otimes {\cal H}$  
when ${\cal H}$ is the module corresponding to the operator
representation given by (\ref{3.3}).

First as a guide we review the route followed successfully and in
generality for the same purpose in the case of the $a_n$
algebras. There one has
\be\label{4.3}
R(u) = u I + \eta P = u I + \eta \left({\textstyle{\frac{1}{2}}} 
\sum_A \lambda_A
\otimes \lambda_A + {\textstyle{\frac{1}{n}}} I\right) \quad ,
\e
using the completeness relation of the Gell-Mann $\lambda$-matrices of
$a_n$. Changing the representation $\frac{1}{2} \lambda_A$ to $X_A$
leads to 
\be\label{4.4}
L(u) = (u \, I + {\textstyle{\frac{1}{n}}}\eta) + \eta \sum_A \lambda_A 
\otimes X_A \quad .
\e
which can be seen to give a `one-site' solution $T(u)=L(u)$ of (\ref{1.2}).

The same approach does not work straightforwardly for $c_2$ (or $c_n$)
because (\ref{4.1}) involves not only $X=x_i \otimes x_i$ but also $Y
= y_a \otimes y_a$, where ({\it cf.}  section 2) the $y_a$ do not belong to
the Lie algebra $c_2$. However, as we have shown above, they can be
expressed in terms of the $x_i$ and so we enforce the change of
representation from $x_i$ to $X_i$ onto $Y$
\be\label{4.5}
y_a \otimes y_a = y_a \otimes {\textstyle{\frac{1}{3}}}
d_{ija} x_i x_j &\mapsto&
y_a \otimes {\textstyle{\frac{1}{3}}} d_{ija} X_i X_j\\
&=& {\textstyle{\frac{1}{6}}} (-\delta_{ij} + \{x_i, x_j\}) \otimes X_i X_j =
{\textstyle{\frac{1}{3}}} (C^2 + 3C -{\textstyle{\frac{1}{2}}} 
{\cal C}^{(2)}) \quad . \nonumber
\e
where, as before, $C = x_i \otimes X_i \equiv x_i X_i \;$, $\;{\cal C}^{(2)}
= X_i X_i$. The results (\ref{2.12}), (\ref{2.11}) and (\ref{2.3})
have all been used here. We do
not know how to handle the complication that follows the general use
of (\ref{4.5}), but we do know how to proceed in special cases, such
as that of section 3, in which ${\cal C}^{(2)} = -\frac{5}{2}$ and
(\ref{3.7}) holds, for then the RHS of (\ref{4.5})
vanishes. Accordingly we seek a solution $T(u)$ or first $L(u)$ of
(\ref{1.2}) in the form
\be\label{4.6}
L(u) = \alpha (u) I + \beta(u) x_i X_i \quad ,
\e
This ansatz succeeds for
\be\label{4.7}
\frac{\alpha(u)}{\beta(u)} = \frac{u-\delta}{\eta}, \quad \delta \in
\R \quad,
\e
and we may put $\beta(u) = \eta$ and  $\alpha(u) = u-\delta$. We emphasise
the non-trivial nature of the result obtained. Since $R(u)$ given by
(\ref{4.1}) necessarily involves $Y=y_a\otimes y_a$ the ansatz
(\ref{4.6}) is not obviously valid a priori in our work. It seems
unlikely to be valid for general representations in the quantum space.
But it works for metaplectic representation of $c_2$,
and likewise for other $c_n$.

\section{Bethe Ansatz for $c_2$ Integrable Models}

Sklyanin's general procedure \cite{skly} for constructing integrable
spin models, {\it e.g.} if operators $X_i=S_i$ acting in ${\cal H}=
{\cal V}_s$ where ${\cal V}_s$
is the vector space in which the spin-$s$ representation of $SU(2)$ acts,
is available here also once we know a basic (`one-site') 
matrix $L(u) \in {\cal V} \otimes {\cal H}$ which
satisfies (\ref{1.2}). We write  
\be\label{5.1}
T(u) = {\cal K} \prod_{r=1}^N L(u-\delta_r) \quad ,
\e
where ${\cal K}$ is a constant matrix such that $[{\cal K}\otimes
{\cal K}, R(u)] =0$. Since $T(u)$ is an $n$-site solution of (\ref{1.2}),
it follows in a well-known way that $[t(u), t(v)] = 0 \;$
where  $t(u) = \mbox{Tr} \; T(u)$, and expansion of $t(u)$ in powers of $u$ 
yields constants of the motion. 

In this section we sketch the solution 
for the eigenstates of $t(u)=\sum_{i=1}^4 T_{ii}$ 
of the system described by (\ref{5.1}) when $L(u)$ is the one-site solution of
(\ref{1.2}) given in the last section
by means of the algebraic Bethe ansatz method \cite{FAD,skly}.
Although detailed attention to ${\cal K}$ and to the $\delta_r$ is important 
for the discussion of the completeness of the set of these eigenstates, we 
defer this to a later publication, here setting ${\cal K}=1$ and  $\delta_r=0$.
We wish rather to explain how the Bethe ansatz method works in our models,
emphasising the simplifications that stem from special features of the 
representation (\ref{3.3}) of $X_i$, and exploiting the
transformation properties of the $T_{ij}$ under the compact $su(2)$ subgroup
of $c_2$. The latter  follow for the  $T_{ij}$ when  ${\bf J}=
\sum_{r=1}^N {\bf J}_r$ is used, $J_r$ for $r=1, \dots N$ being defined 
at the $r$-th
site in terms of the oscillator variables of that site as in (\ref{3.3}).
Perhaps it should be pointed out that ${\bf J}$ as just defined is not 
equal to the $(N-1)$-fold coproduct of the 
`one-site' operator $(J_+=T_{21}^{(1)} \; , \;
J_z=\frac{1}{2}(T_{11}^{(1)}-T_{22}^{(1)}) \; , \; J_-=T_{12}^{(1)} \, )$, in
the sense of the Yang-Baxter algebra (\ref{1.2}). 
Nevertheless its commutation relations
with the $\Delta^{(N)} T_{ij}^{(1)}$, involving the
standard comultiplication, can be shown explicitly to give these operators
the correct $su(2)$ tensor operator properties.

Our sketch deals only with the tower of Bethe states based in the Fock 
vacuum  of both the oscillators at each of the $N$ sites. Similar work for
other towers whose ground states involve oscillator states of occupation number
one requires only a modest extension of the analysis described here. See
\cite{MPW}. We note first, from (\ref{3.3}), that all the $L_{ij}$ with $i<j$,
as well as $L_{21}$ and $L_{43}$ annihilate
the Fock vacuum at each site, and it is 
easy to promote the same result to the same set of $N$-site $T_{ij}$ of
(\ref{5.1}). Also, at any site we have $(L_{42}-L_{31}) | 0 \rangle =0$, and
this result to can be similarly promoted to
\be\label{5.2}
(T_{42}-T_{31})(u) |0 \rangle =0 \quad .
\e
We aim first to construct the Bethe states
\be\label{5.3}
T_{41}(v) |0 \rangle \quad ,\quad  T_{42}(v_1) T_{31}(v_2) |0 \rangle 
\quad , \quad \dots \quad ,
\e
which have $j=m=1,2, \; \dots\; $ 
quantum numbers w.r.t ${\bf J}$. Then the other
states with the same $j$ and lower $m$ follow by application of the lowering
operator $J_-$. Since $[{\bf J}, t(u)]=0$ all the states of each multiplet
of Bethe states have the same eigenvalue of $t(u)$, and allowed $v$-values 
given by the same set of Bethe equations. One naturally expects (and finds)
that there are, alongside the $j=2$ multiplet, $j=1$ and $j=0$ multiplets
bilinear in the creation operators $T_{41},T_{31}, T_{42}$ and $T_{32}$. There 
is not scope in this paper to describe in full the subtleties of the analysis
(in which (\ref{5.2}) allows significant simplifications), so next we 
present results.

Let $w=u-v$, let $\tau_+(u)$ denote the eigenvalue of $T_{11}$ and of $T_{22}$
for the vacuum state $| 0 \rangle$, and let $\tau_-(u)$ do the same for
$T_{33}$ and $T_{44}$. It is easy to calculate $\tau_{\pm}(u)$ using 
(\ref{5.1}) and (\ref{3.3}).
Then for  $j=1$, we find
\be\label{5.4}
t(u) T_{41}(v) |0 \rangle =\tau(u) T_{41}(v) |0 \rangle \quad , 
\e
where the eigenvalue $\tau(u)$ is given by
\be\label{5.5}
\tau(u)=(1+{\textstyle{\frac{\eta}{w}}}) \tau_+
(u)+(1-{\textstyle{\frac{\eta}{w}}}) \tau_-(u) \quad ,
\e
and $v$ is determined by the Bethe equations
$\tau_+(v)=\tau_-(v) \quad $.

For  $j=m=2$, the eigenvalue $\tau(u)$ of $t(u)$ for 
$T_{41}(v_1) T_{41}(v_2) |0 \rangle$, and the Bethe equations for $v_1$
and $v_2$ are given by
\begin{eqnarray}\label{5.6}
\tau(u) & = & 2(1+{\textstyle{\frac{\eta}{w_1}}}
+ {\textstyle{\frac{\eta}{w_2}}} +2{\textstyle{\frac{\eta^2}{{w_1}{w_2}}}}) 
\tau_+(u)
\; \nonumber \\
     & + & 2(1-{\textstyle{\frac{\eta}{w_1}}}- {\textstyle{\frac{\eta}{w_2}}} +
2{\textstyle{\frac{\eta^2}{{w_1}{w_2}}}}) \tau_-(u) 
 \; ,  
\e where we write $w_k=u-v_k$, $k=1,2$ and $v=v_1-v_2$, and
\be\label{5.7}
\tau_+(v_1) / \tau_-(v_1) =\tau_-(v_2) / \tau_+(v_2)=(v+2\eta)/(v-2\eta)
\quad .
\e
The generalisation to higher integral $j$-values is evident. Turning to $m=1$
states bilinear in the creation $T_{ij}$, we find two, one that follows by 
application of $J_-$ to the state just discussed. Writing
\begin{eqnarray}\label{5.8}
 \phi_1 & = & T_{42}(v_1) T_{41}(v_2) |0 \rangle \; \quad , \nonumber \\
 \phi_2 & = & T_{31}(v_1) T_{41}(v_2) |0 \rangle \; \quad , \nonumber \\
 \phi_3 & = & T_{41}(v_1) T_{42}(v_2) |0 \rangle \; \quad , \nonumber \\
 \phi_4 & = & T_{41}(v_1) T_{31}(v_2) |0 \rangle \; \quad ,  
\end{eqnarray} 
we find the $|21\rangle $ 
state is given by $(\phi_1+\phi_2+\phi_3+\phi_4)$, and the unique 
$|11\rangle $ Bethe state in the context is given by
$\{ (\phi_1+\phi_2-\phi_3-\phi_4) +
{\textstyle{\frac{2\eta}{v}}} (\phi_1-\phi_2)\}$ 
with $t(u)$ eigenvalue given by $2(1+\frac{\eta}{w_1}+ \frac{\eta}{w_2}) 
\tau_+(u)+
2(1-\frac{\eta}{w_1}- {\frac{\eta}{w_2}}) \tau_-(u) $, 
and $v_1$ and $v_2$ determined by the Bethe 
equations $\tau_+(v_1)=\tau_-(v_1) \; , \tau_+(v_2)=\tau_-(v_2)$. The
pattern of the Bethe states should by now be clear.

The states constructed have various nice properties. Their Bethe equations
ensure that the poles of their $t(u)$ eigenvalues have residues zero. They are
also invariant under the exchange of $v_1$ and $v_2$, although proof of this 
is not obvious, requiring detailed use of the commutation relations
of the $T_{ij}$ that we are using.
 
A brief sketch of how these results are derived is called for, because there
is a level of complication not seen in Bethe ansatz studies in models with
$a_1=su(2)$ invariance. This is no doubt the price to be paid for obtaining 
results without recourse to a nesting process. Complications not present 
in $a_1$ studies come into play here whenever the third term of 
(\ref{4.1}) enters the commutation  relations amongst the $T_{ij}$ 
non-trivially. A sufficient illustration of what has to be done whenever this
happens is provided by showing how to compute the effect of the term 
$T_{11}(u)$ of $t(u)$ on $T_{41}(v) | 0 \rangle $. We here shall
abbreviate any $T_{ij}(u)$ or $T_{ij}(v)$ by $T_{ij}$ or $T_{ij}^{\prime}$.
Our strategy is push to the right end of any term only factors that annihilate 
$| 0 \rangle$, or else are of the type $T_{kk}$, with no sum on $k$, which 
multiply $| 0 \rangle$ by $\tau_{\pm}$. Direct calculation of 
$w T_{11} T_{41}^{\prime}$ from (\ref{1.2})  
fails to achieve this, leaving a term in $T_{21} T_{31}^{\prime}$ 
that has to be 
eliminated by calculation from (\ref{1.2}) of $w T_{21} T_{31}^{\prime}$.
This two stage procedure yields finally the result
\be\label{5.9}
T_{11} T_{41}^{\prime}=(1+\frac{2\eta}{w}) T_{41}^{\prime} T_{11} -
\frac{2\eta}{w} 
T_{41} T_{11}^{\prime} +\frac{\eta}{w}  (T_{31}^{\prime} T_{21}-T_{31} 
T_{21}^{\prime}) \quad .
\e Another example yields
\begin{eqnarray}\label{5.10}
T_{11} T_{42}^{\prime} & = & (1+\frac{\eta}{w+\eta})T_{42}^{\prime} T_{11}
+(\frac{2\eta}{w}-\frac{\eta}{w+\eta}) T_{41}^{\prime} T_{12}-\frac{2\eta}{w}
T_{41} T_{12}^{\prime}  \nonumber \\
& - &  \frac{\eta}{w} T_{31} T_{22}^{\prime}+(\frac{\eta}{w}-
\frac{\eta}{w+\eta}) T_{31}^{\prime} T_{22}+\frac{\eta}{w+\eta}
T_{32}^{\prime} T_{21} \quad .  
\end{eqnarray}
One must push each of the four pieces of $t(u)$ past each of the creator 
$T_{ij}$, needing a total of about 32 results in all, to deduce results 
such as those just quoted. This entails eight results like each of
(\ref{5.9}) and (\ref{5.10}); the remainder do not involve third term of 
(\ref{4.1}) and are written down directly.

\section{Discussion}

We have described, for $c_2$, formalism and the construction of a family of
integrable models that generalise in every respect to all $c_n$. This
applies to the construction (\ref{3.2}) and the proof that the $X_i$
obey the same commutation relations as the $x_i$, to the deduction of
relations (\ref{4.6}) and (\ref{4.7}). A simliar construction is
available for the  $d_n$ series, where we use fermionic
instead of bosonic oscillator variables.
Note that in the $c_n$ case, there is a maximal compact $u(n)$ subalgebra,
and sets of creation operators $T_{ij}$ which transform according to its 
dimension $n$
defining representation $(1, \; \dots \; ,0)$, leading to a $u(n)$ multiplet 
structure  of Bethe eigenstates \cite{MPW}, with a pattern similar to that
indicated in section five.
We note also that our method of 
getting Bethe eigenstates without recourse to nesting applies equally well
to the treatment \cite{MPW} of $c_n$ Gaudin models \cite{Gau}.

{\bf Acknowledgement.} This work was partly supported by PPARC. F.W. is 
grateful to Trinity College, Cambridge for an IGS, and to EPSRC 
for a research grant. We thank Hendryk Pfeiffer for useful discussions.


\begin{thebibliography}{unsrt}

\bibitem{IKMT} A.G. \ Izergin, N. \ Kitanine, J.M. \ Maillet and V. \ Terras.
{\sl Spontaneous magnetization of the $XXZ$ Heisenberg spin chain}. Preprint
Lyon, France, LPENSL-TH-13-98.

\bibitem{FAD} L.D. \ Fadde'ev.
{\sl Recent advances in field theory and
statistical mechanics}. Les Houches Lectures, eds. J.B. \ Zuber and R. \ Stora,
Elsevier,1984, pp561-608.
 
\bibitem{jimbo} M. \ Jimbo. {\sl Introduction to the Yang-Baxter equation}.
Internat. J. Phys. {\bf 4} 3759-3777 (1989).

\bibitem{skly} E.K. \ Sklyanin. {\sl Separation of variables: new trends}.
Prog. Theor. Phys. Supp.. {\bf 118} 35-60 (1995).

\bibitem{GraS} C. \ Gomez, M. \ Ruiz-Altaba and G. \ Sierra. {\sl Quantum 
Groups in two dimensional physics}. Cambridge University Press, 1996.

\bibitem{Resh1} N. Yu. \ Reshitikhin. {\sl Calculation of the norms of Bethe
Vectors for $SU(3)$ symmetry}. Zap. Nauchnyk. Sem. LOMI {\bf 150} 196-213
(1986) or J. Sov. Math. {\bf 46} 1694-1706 (1989).

\bibitem{Resh2} N. Yu. \ Reshitikhin. {\sl Algebraic Bethe ansatz for $SO(N)$
invariant transfer matrices}. Zap. Nauchnyk. Sem. LOMI {\bf 169} 128-140
(1988) or J. Sov. Math. {\bf 54} 940-951 (1991).

\bibitem{Resh3} N. Yu. \ Reshitikhin. {\sl Integrable models of quantum
magnets}. TMF {\bf 63} 347-366 (1985) or Theor. Math. Phys. {\bf 63}
555-569 (1985).

\bibitem{deV} O. \ Babelon, H.J. \ de Vega and C.M. \ Viallet. {\sl Exact
solution of the $Z_{n+1} \otimes Z_{n+1}$ symmetric generalisation of the $XXZ$
model}. Nucl. Phys. {\bf B200} 266-280 (1982).

\bibitem{FFR} B. \ Feigin, E. \ Frenkel and N. Yu. \ Reshitikhin.
{\sl Gaudin model, Bethe ansatz and critical level}. Commun. Math. Phys.
{\bf 166} 27-62 (1994).

\bibitem{BABfl} H.M. \ Babujian and R. \ Flume. {\sl Off-shell Bethe ansatz
equations and
$N$-point conformal correlators in the $SU(2)$ WZNW theory}. Mod. Phys. 
Lett. {\bf A9} 2029-2040 (1994).

\bibitem{MPW} A.J. \ Macfarlane, Hendryk Pfeiffer and F. \ Wagner. DAMTP
Cambridge preprints, in preparation.

\bibitem{Itz} C. \ Itzykson. {\sl Remarks on the bosonic commutation rules}.
Commun. Math. Phys. {\bf 4} 92-122 (1967).

\bibitem{SW} S. \ Sternberg and J.A. \ Wolf. {\sl Hermitian Lie algebras and 
metaplectic representations}. Trans. Am. Math. Soc. {\bf 238} 1-43 (1976).

\bibitem{KS} P.P. \ Kulish and E.K. \ Sklyanin. {\sl Solutions of the 
Yang-Baxter equation}.  Zap. Nauchnyk. Sem. LOMI {\bf 95} 129-160
(1980) or J. Sov. Math. {\bf 19} 1596-1620 (1982).

\bibitem{Gau} M. \ Gaudin. {\sl Diagonalisation d'une classe d'Hamiltonians 
de spin}. J. Physique {\bf 37} 1087-1098 (1976); {\sl La fonction d'onde de 
Bethe}. Masson 1983.

\end{thebibliography}
\end{document}